\documentclass[pra,reprint,showpacs]{revtex4-1}

\usepackage{epsfig}
\usepackage{graphics}
\usepackage{amsmath}
\usepackage{color}

\newcommand{\ket}[1]{\left\vert#1\right\rangle}
\newcommand{\bra}[1]{\left\langle#1\right\vert}

\newcommand{\eqd}{\buildrel D \over =}

\def\be{\begin{equation} }
\def\ee{\end{equation} }
\begin{document}

\author{Yoav Sagi, Rami Pugatch, Ido Almog, Nir Davidson}
\affiliation{Department of Physics of Complex Systems, Weizmann Institute of Science, Rehovot 76100, Israel}
\author{and Michael Aizenman}
\affiliation{Departments of Physics and Mathematics, Princeton University, Princeton NJ 08544, USA}

\title{Motional Broadening in Ensembles With Heavy-Tail Frequency Distribution}

\begin{abstract}
We show that the spectrum of an ensemble of two-level systems can be broadened through `resetting' discrete fluctuations, in contrast to the well-known motional-narrowing effect. We establish that the condition for the onset of motional broadening is that the ensemble frequency distribution has heavy tails with a diverging first moment. We find that the asymptotic motional-broadened lineshape is a Lorentzian, and derive an expression for its width. We explain why motional broadening persists up to some fluctuation rate, even when there is a physical upper cutoff to the frequency distribution.
\end{abstract}
\maketitle

\section{Introduction}
Ensembles consisting of many two-level system (TLS) are of interest in many physical disciplines.  In precision metrology, the energy difference between such two levels in a cesium atom is used to define the second. In quantum computation, the TLS is called a qubit and it replaces the classical bit with an added ability to store any superposition of the two underlying logical values. An ensemble of such TLS can serve as a storage medium for quantum information \cite{Fleischhauer2002}. A quantum memory is a necessary building block in quantum networks since it relieves the requirement for a series of successful quantum operations and therefore enable scalability \cite{Duan2001,Kimble2008}. In order to extend the coherence time of such a memory it is first essential to understand the effect of its coupling to the environment.

The difference in the energy levels, or the transition frequency between the two,  is in general not identically the same for all the TLS in the ensemble, and its distribution has a certain breadth.   For instance, for a cold atom ensemble held together by a confining potential, the difference between the energy levels of the internal states, may depend on the location of the atom within the trap, and may also be affected by inter-particle interactions. Although the ``bare spectrum'' due to the inhomogeneities within the ensemble is usually time-independent, the frequencies of the individual TLS are in general not constant, and undergo time-dependent fluctuations.

Such fluctuations are usually thought to induce narrowing of the power spectrum in observations of the decay rate of the ensemble coherence, a phenomenon named motional narrowing.  Historically, motional narrowing was first observed in NMR, where the spectra of liquid materials were found to be significantly narrower than those of solids due to thermal motion of the nuclei \cite{PhysRev.73.679}. Spectral narrowing due to fluctuations was later encountered in many other fields including in hot atomic vapors (Dicke narrowing) \cite{PhysRev.89.472}, semiconductor microcavities \cite{PhysRevLett.77.4792}, quantum dots \cite{Berthelot2006} and trapped cold atomic ensembles \cite{PhysRevLett.105.093001}.

In this paper we show that fluctuations can have the reverse effect and lead to broadening of the spectrum (motional broadening). In terms of quantum information, motional broadening manifests itself as a shortening of the coherence time as the the fluctuation rate increases. We prove that the condition for the emergence of motional broadening is that the ensemble frequency distribution will have heavy tails with a diverging mean. An example for this effect was first pointed out in Ref. \cite{Burnstein1981335}. We also show that for both motional narrowing and broadening the asymptotic decay of the coherence is exponential, and derive an expression for the decay rate. Since in practice heavy tails of the frequency distribution can be sustained only up to some point, we study scenarios with cutoffs and show that motional broadening persists up to some fluctuation rate. The motional broadening phenomenon should be relevant to many fields in which heavy-tail distributions are encountered, including turbulence \cite{RevModPhys.73.913}, diffusion \cite{Bouchaud1990127} and laser-cooling \cite{levy_statistics_and_laser_cooling_book}.

This paper is organized as follows. In section \ref{the_model_section} we describe the model and the quantity which is measured in the experiment, i.e. the coherence function. To show that the effect of the increase in the fluctuation rate (in time) depends on the tail of the probability distribution, we shall first consider it within the context of the so-called `stable laws', whose defining property is naturally suited for our purpose (section~\ref{stable_distributions_section}). The rule is then shown to apply also beyond those special cases by first demonstrating it on a family of probability distributions with a tunable parameter (the `student's t-distribution', section~\ref{sec:studentsT}). The main result is then generalized to distributions which converge to stable distributions (section \ref{Generalization_section}).

Further discussion is presented in section \ref{Discussion_section}: we note that motional broadening is analogous to the anti-Zeno effect, and draw an analogy between the spectroscopic problem we have considered and diffusion in real space. We use this analogy to propose an experiment to observe the phenomenon. We continue with the discussion of the effect of a cutoff in the distribution, and the asymptotic shape and width of the broadened spectrum. Our conclusions are given in section \ref{conclusions_section}.

\section{The model}\label{the_model_section}

We consider an ensemble of two-level systems (TLS) with internal states designated by $\ket{1}$ and $\ket{2}$, which are eigenstates of the Hamiltonian.  Their  energies are of the form
\begin{equation}
 E_{1,2} \ = \  E_0 \pm \hbar (\omega_0 + \delta(t))/2 \,
\end{equation}
with $\hbar\omega_0$ the `base line' energy difference, and $\hbar\delta(t)$ the detuning from this base line. $\delta(t)$ is a random correction whose value  fluctuates  within the ensemble and which for individual TLS  also changes discontinuously in time.
We assume a steady-state situation in which the distribution of the `detuning' term $\delta(t)$ is time-independent, and denote by $P_0(\delta) d \delta$ its  distribution  over the ensemble.

In a Ramsey experiment - like  procedure, e.g., as described in~\cite{PhysRevLett.105.093001}, a $\pi/2$ pulse may be used to initiate the ensemble's individual TLS in the coherent superposition states
$\ket{\psi} = 1/\sqrt{2} (\ket{1} + \ket{2})$.   In a situation where the internal eigenstates can be treated as stationary, but the energy differences between them fluctuate in time the individual TLS states' density matrices evolve into
\begin{multline}
\ket{\psi(T)} \bra{\psi(t)} = \frac{1}{2} [ \ket{1} \bra{1}  +  \ket{2} \bra{2}] \
\\
 + \frac{1}{2}  e^{-i[\omega_0 T+\phi(T)]/2} \ket{1}\bra{2}
   + \frac{1}{2}  \, \, e^{ i[\omega_0 T+\phi(T)]/2} \ket{2}\bra{1}   \, .
\end{multline}
where the fluctuating part of the accumulated phase difference between the two internal states
is
\be
\phi(T)=  \int_0^T \delta(t) dt  \, .
\ee

The phase difference in the off-diagonal terms can be determined using a second  $\pi/2$ pulse, applied at time $t$.  The measured quantity  is described by the coherence function:
\be
R(T)\  =\   \left |\langle e^{i  \phi(T) }  \rangle  \right|  \, ,
\ee
 where   $\langle ... \rangle$ denotes the ensemble average \cite{cywinski:174509}.

Our discussion concerns the effect on the  ensemble coherence function  of the rate at which the individual detuning terms $\delta(t)$ are refreshed.  In a model which may capture the effect of `hard collisions' these terms change discontinuously at random times, $\{t_j\}$, at which the value of  $\delta(t)$ is reset with the stationary distribution $P_0(\delta) d \delta$.

The coherence function starts at $R(0) = 1$ and decays to $0$ on a time scale  which is referred to as the \emph{coherence time}. If the absolute value is omitted from its definition, oscillations may occur.
 The spectrum function, $S(\omega)$, is the absolute value squared of the coherence function's Fourier transform.   In general, the width of the spectral distribution varies in the opposite way to the length of the coherence time, e.g, spectral narrowing corresponds to the extension of the coherence time.

As long as the fluctuations in time are negligible, the detuning of each TLS occurs at a constant rate and the ensemble coherence  is given by
\be R_0(T) \ =\  \left |\int_{-\infty}^{\infty}P_0(\delta)e^{i\delta T}d\delta  \right| \, .
\ee

It was already noted that under certain conditions the coherence time is \emph{increased} due to fluctuations (motional narrowing). The main result presented here is that the coherence time can also be \emph{reduced} by the increase of the fluctuation rate. The latter happens when the detuning distribution $P_0(\delta)$ has heavy tails, with a diverging first moment (to avoid confusion, we shall henceforth reserve the term fluctuations for \emph{fluctuations in time} of individual TLS, not to the \emph{statistical fluctuations} which occur within the ensemble).

\section{Motional broadening for stable distributions}\label{stable_distributions_section}

It is instructive to consider the effect of fluctuations when the distribution of the detuning $\delta$ is given by one of the so-called `stable laws' with a  characteristic exponent $0<\alpha\leq 2$~\cite{feller_book_specific_1}.
Some well-known examples of stable distributions are Gaussian ($\alpha=2$), Cauchy ($\alpha=1$) and L\'{e}vy ($\alpha=1/2$) distributions.
For each distribution in this class, the weighted sum of independent identically distributed variables produces a variable with a scaled version of the same distribution. More explicitly: for any two real numbers $q,s>0$, and a pair of independent variables $\delta_1,\delta_2$ of such distribution, the weighted sum $\left(q\delta_1+s \delta_2\right)$ has the same distribution as $(q^{\alpha}+s^{\alpha})^{1/\alpha} \, \delta$.   The characteristic function of any $\alpha$-stable distribution satisfies~\cite{feller_book_specific_1}:
\be  \label{eq:alpha_phi}
|\varphi_\phi(t)| \equiv \left| \langle   e^{-it \delta } \rangle \right| \ = \  e^{-c_\alpha |t|^\alpha} \, ,
\ee
with some $c_\alpha>0$

Assuming $P_0$ is $\alpha$-stable, the accumulated phase due to the discrete fluctuation events  (`collisions') at which $\delta(t)$ is reset can be written as
\begin{equation}
\phi(T)=\sum_{j=1}^n \Delta t_j \delta_j \eqd \left(\sum_{j=1}^n \tau_j^\alpha \right)^{1/\alpha} \delta T\ \ ,
\end{equation}
with $\Delta t_j$ denoting the periods between collisions, $\tau_j = \Delta t_j /T$, and $\eqd$ standing for the equivalence of the corresponding distributions.  By \eqref{eq:alpha_phi} the coherence without collisions is given by $R_0(T)=e^{-c_\alpha T^\alpha}$ and with collisions it is given by
\begin{equation}\label{coherence_stable_with_col}
R(T) \  = \   e^{-c_\alpha (\sum_{j=1}^n \Delta t_j^\alpha)} \ = \  R_0(T)^{\sum_{j=1}^n \tau_j^\alpha}\ \ .
\end{equation}
For any series of collisions:  $\sum_{j=1}^n \tau_j=1$, and hence:
\begin{equation}\label{deltaT}
 \sum_{j=1}^n \tau_j^\alpha   \  = \
 \sum_{j=1}^n  \tau_j \ \tau_j^{(\alpha-1)}   \
\begin{cases}
\le  (\tau_{max})^{\alpha-1}  & \alpha\geq 1  \\[2ex]
\ge  (\tau_{max})^{\alpha-1}  & \alpha\leq 1
\end{cases}
\end{equation}
with $\tau_{max} =\max \{ \tau_j\}$.

Combining \eqref{coherence_stable_with_col}, \eqref{deltaT} and the fact that $\tau_{max}\le 1$ we find that the
transition from motional narrowing to broadening occurs at $\alpha=1$:
\begin{equation}\label{main_result_for_stable_distributions}
R(T)\begin{cases}
> R_0(T)\,,   & \alpha > 1 \ \ \text{(motional narrowing)}\\
<  R_0(T) \,, & \alpha< 1 \ \ \text{(motional broadening)} .
\end{cases}
\end{equation}
It may be noted that:
\begin{enumerate}
\item The above holds regardless of   the distribution, or the values, of the collision times
\item  If the time between collisions is constant,  $\Delta t_j = t_{coll}$, then the coherence function decays exponentially in $T$:
\be
R(T)  \ = \  e^{-c_\alpha (t_{coll})^{(\alpha -1)}  \, T}\, .
\ee
That is: the coherence time is given by $c_\alpha (t_{coll})^{(\alpha -1)}$, which is in
 line with the observation made in~\eqref{main_result_for_stable_distributions}.  A similar conclusion holds if $\Delta t_j $ are random but of common  order $t_{coll}$.
\end{enumerate}

\section{The solvable case of Poisson fluctuations}\label{Poisson_fluctuations_section}
An explicit expression for $R(t)$ can be derived for discrete fluctuations (e.g., collisions) which occur with a Poisson distribution in time, at  rate $\Gamma$. After a randomizing event, the TLS acquires a new detuning rate, independently drawn with the probability distribution $P_0(\delta)$.  The coherence function can be calculated exactly in this model through the following expression for its Laplace transform,
$\widetilde{R}(s) = \int_0^\infty e^{-sT} R(T) dT$,~\cite{brissaud:524,PhysRevLett.104.253003}:
\begin{equation}\label{equation_of_spectrum}
\widetilde{R}(s)=\frac{\widetilde{R}_0(s+\Gamma)}{1-\Gamma\widetilde{R}_0(s+\Gamma)} \ \ .
\end{equation}
This relation is valid for all $P_0(\delta)$. It was shown experimentally that  \eqref{equation_of_spectrum} correctly describes the effect of cold-collisions in trapped atomic ensembles \cite{PhysRevLett.104.253003}.
Equation~\eqref{equation_of_spectrum} is also useful for model calculations, such as the one  presented next.

\section{An illustrative example}  \label{sec:studentsT}

\begin{figure}
    \begin{center}
    \includegraphics[width=8cm]{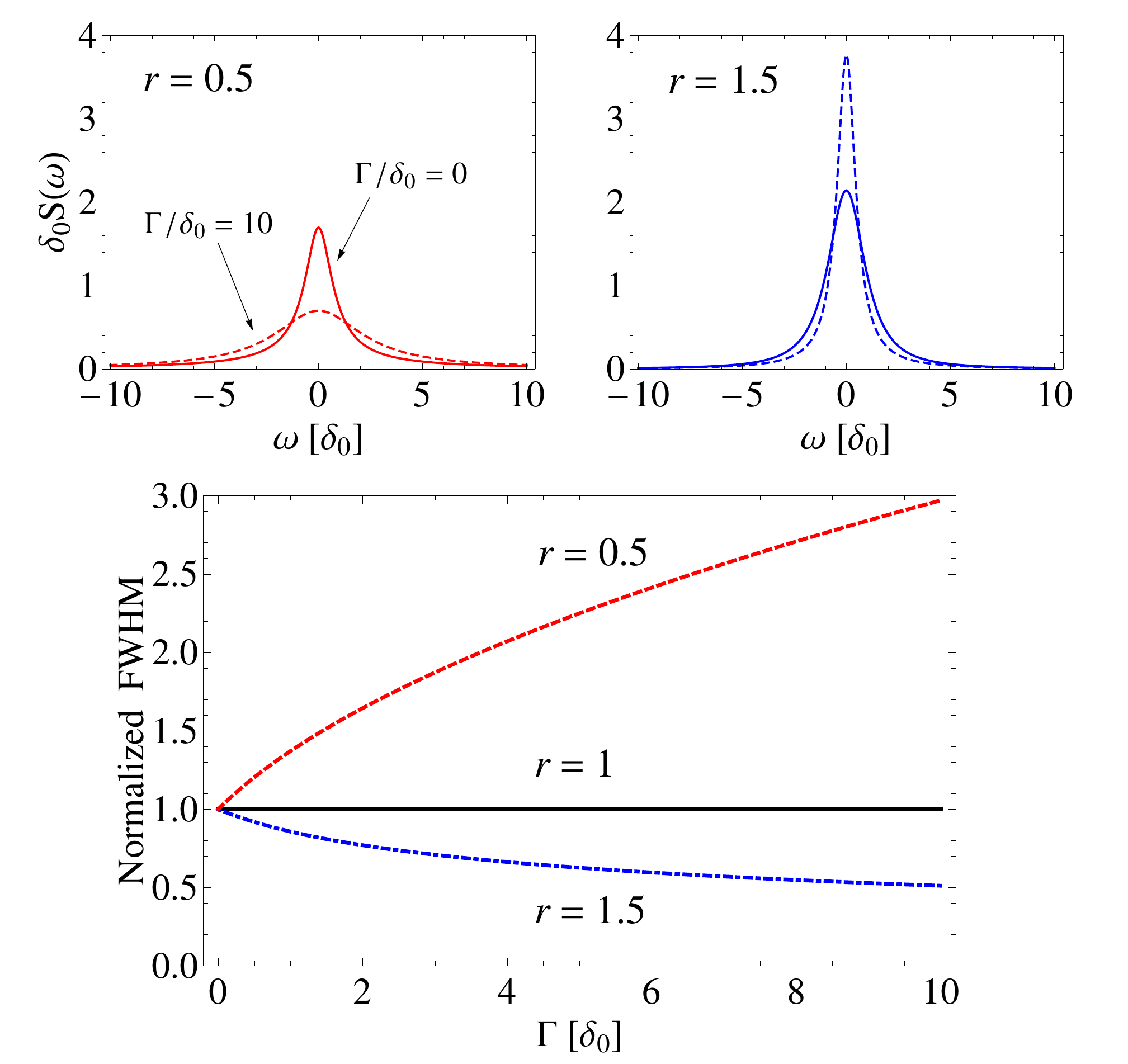}
    \end{center}\caption{The two upper graphs show the spectrum for a Student's t-distribution of the detunings [see \eqref{students_t_distribution}]. On the left the spectrum is plotted for $r=0.5$ for which the first moment of the distribution diverges whereas on the right the spectrum is plotted for $r=1.5$ for which only the second moment diverges but the first moment exists. For both spectrums the spectrum is plotted without fluctuations (solid line) and with fluctuations at $\Gamma=10\delta_0$ (dashed line). In the lower graph we plot the full width at half the maximum (FWHM) normalized to the FWHM without fluctuations, as a function of $\Gamma$.}\label{FWHM_vs_Gamma_graph}
\end{figure}

To illustrate the transition from motional narrowing to broadening beyond the above class of special  distributions, we consider an ensemble with a Student's t-distribution, with a tunable parameter $r$:
\begin{equation}\label{students_t_distribution}
P_0(\delta)=N(r,\delta_0)\left[ 1+\frac{1}{r}\left(\frac{\delta}{\delta_0}\right)^2\right]^{-\frac{1+r}{2}} \ \ ,
\end{equation}
with the normalization factor
\begin{equation}
N(r,\delta_0)=\Gamma\left(\frac{r+1}{2}\right)/\Gamma\left(\frac{r}{2}\right)\delta_0\sqrt{r\pi} \ \ ,
\end{equation}
where $\Gamma(z)$ is the gamma function. For $r\rightarrow\infty$ the distribution is approaching a Gaussian with a standard deviation $\delta_0$, and for $r=1$ it is identical to the Cauchy distribution (Lorentzian). The first (second) moment of the distribution diverges for $r<1$ ($r<2$). Using \eqref{equation_of_spectrum}, we calculate the spectrum and plot it in Fig. \ref{FWHM_vs_Gamma_graph} for $r=0.5$ and $r=1.5$, with and without fluctuations. For $r=1.5$ we observe that the spectrum becomes narrower in the presence of fluctuations, which demonstrate that motional narrowing persists even when the second moment diverges. On the other hand, for $r=0.5$ the fluctuations broaden the spectrum. In Fig. \ref{FWHM_vs_Gamma_graph} we plot as a function of $\Gamma$ the normalized spectral width, which is defined to be the full width at half the maximum (FWHM) divided by the FWHM for $\Gamma=0$. The figure clearly shows the narrowing (for $r=1.5$) or broadening (for $r=0.5$) effects as the fluctuation rate increases. Curiously, for the Cauchy distribution, corresponding to $r=1$, there is no $\Gamma$ dependency. This fact is in line with both \eqref{main_result_for_stable_distributions} and \eqref{equation_of_spectrum}.

\section{Generalization}\label{Generalization_section}
The above observations can be extended further to distributions which are by themselves not stable, but are in the domain of attraction of a stable distribution (or `law') $S_\alpha$.  This notion means that a sum of variables drawn from the distribution, up to a normalizing factor, converges in distribution to $S_\alpha$, as the number of summands increases.

A distribution belongs to the domain of attraction of an $\alpha$-stable law if its cumulative distribution function, $F(x)$, scales as $F(x)\sim |x|^\alpha h(|x|)$ as $x\rightarrow -\infty$, and $1-F(x)\sim x^\alpha h(x)$ as $x\rightarrow \infty$, where $h(x)$ is a slowly varying function at infinity \cite{Ibragimov_and_Linnik_book}. The domain of attraction of the Gaussian distribution contain all distributions with a finite variance.

If $P_0$ belongs to the domain of attraction of an $\alpha$- stable distribution, then its characteristic function is of the form $|\varphi_\phi(t)|=e^{-c|t|^\alpha\tilde{h}(t)}$, with $\tilde{h}(t) =e^{o(t)}$ a slowly varying function as $t\rightarrow 0$ \cite{Ibragimov_and_Linnik_book}. The coherence without collisions is given by $R_0(T)=\exp\left[-c T^\alpha\tilde{h}(T)\right]$, and with collisions it is given by $R(T)=\exp\left[-c \sum_{j=1}^n \Delta t_j^\alpha\tilde{h}(\Delta t_j)\right]$. Extending \eqref{deltaT} one may see that if $\lim_{n\rightarrow\infty} \tau_{max}=0$, at a fixed $T$, than for $\alpha>1$: \mbox{$\sum_{j=1}^n \Delta t_j^\alpha\tilde{h}(\Delta t_j)<C_1 T^\alpha \tau_{max}^{\eta_1}$}, and for $\alpha<1$: \mbox{$\sum_{j=1}^n \Delta t_j^\alpha\tilde{h}(\Delta t_j)> C_2 T^\alpha \tau_{max}^{-\eta_2}$}, with some $\eta_j,C_j>0$. This extends the validity of \eqref{main_result_for_stable_distributions} for the limit of many randomizing events (high collision rate) when the detuning have a distribution in the domain of attraction of an $\alpha$-stable law.

\section{Discussion}\label{Discussion_section}
The above can be summarized by saying that whether the coherence of a TLS ensemble decays faster or slower due to `resetting' discrete fluctuations depends on the the tails of the detunings distribution. Motional broadening emerges for heavy-tailed distributions with $\alpha<1$ which corresponds to a diverging first moment. This explains the results of Fig. \ref{FWHM_vs_Gamma_graph} since the Student's t-distribution belongs to the domain of attraction of an $\alpha$-stable distribution with $\alpha=r$. Furthermore, the criterion for motional broadening given in \eqref{main_result_for_stable_distributions} coincides with the divergence of the detuning distribution's first moment.

\begin{figure}
    \begin{center}
    \includegraphics[width=8cm]{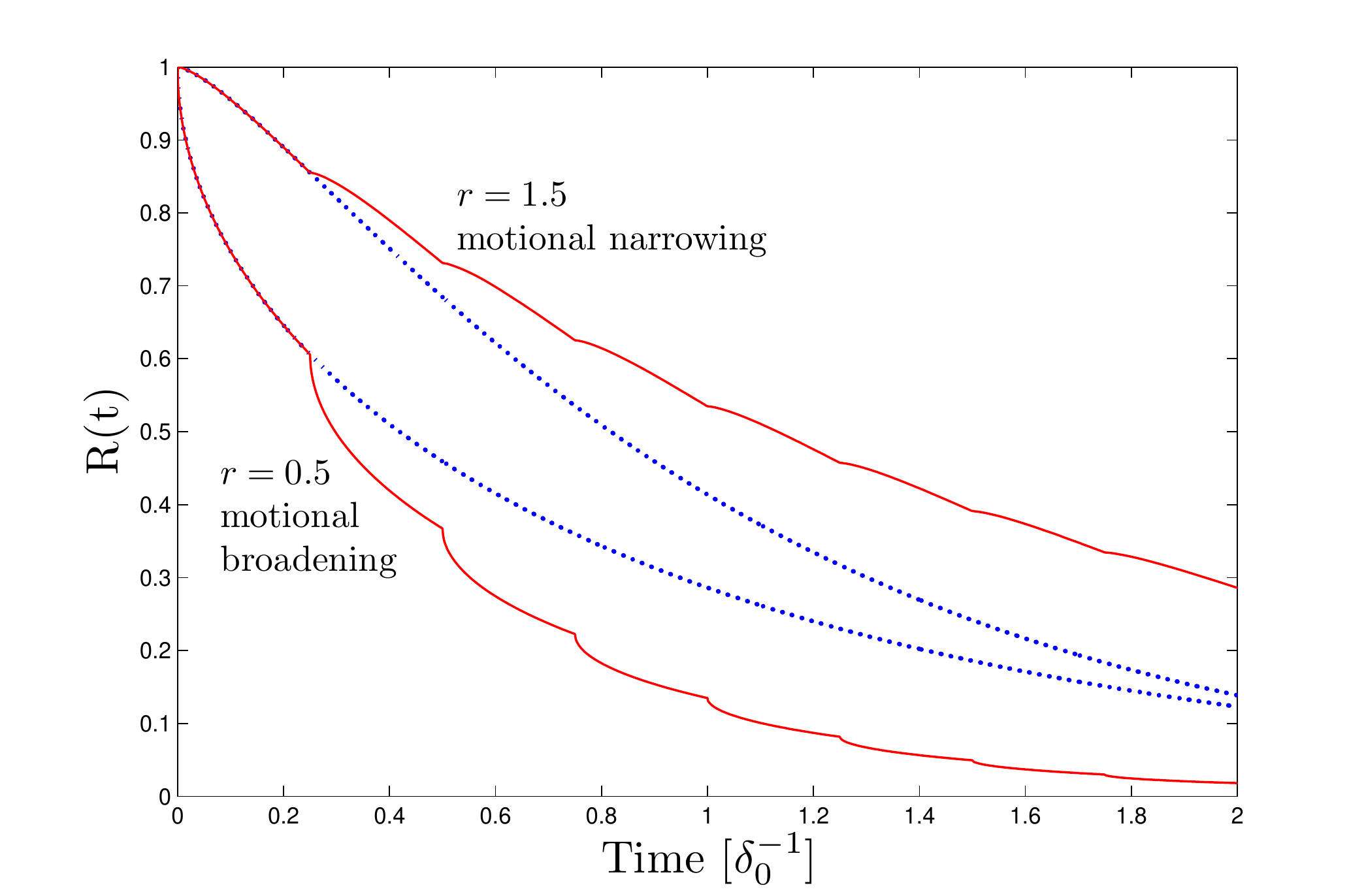}
    \end{center}\caption{The coherence without fluctuations (dotted lines) and with fluctuations separated by $\Delta t_j=0.25 \cdot\delta_0^{-1}$ (solid lines). The detuning distribution is assumed to be the Student's t-distribution with $r=0.5$ and $1.5$. Since the coherence is given by $R(T)=R_0(\Delta t_1)\cdot R_0(\Delta t_2)\cdots R_0(\Delta  t_n)$, a Zeno/anti-Zeno like behavior explains the transition from motional narrowing to broadening for a diverging first moment of $P_0$, at which point $\partial_T R_0|_{T=0}$ changes from $0$ to $-\infty$.}\label{zeno_like_illustration}
\end{figure}

\emph{The relation to the anti-Zeno effect}.---To get an intuition for the above results we write the coherence at a time $T$, for a given series of randomization events, as
\begin{eqnarray}
R(T)&=&\prod_{l=1}^{n}\int_{-\infty}^{\infty}d\delta_l P_0(\delta_l)e^{i\Delta t_l \delta_l} \nonumber\\
&=&R_0(\Delta t_1)\cdot R_0(\Delta t_2)\cdots R_0(\Delta t_n) \ \ .
\end{eqnarray}
In addition, the derivative $\partial_T R_0|_{T=0^+}$ is $0$ for $\alpha>1$ and $-\infty$ for $\alpha<1$. The combination of these two properties explains why depending on whether $\alpha$ is larger or smaller than $1$, the coherence after a resetting collision lie above or below the original curve of $R_0(T)$, as depicted in Fig. \ref{zeno_like_illustration} for the simple case of equal times between such collisions. In this respect, motional narrowing is analogous to the Zeno effect \cite{Milburn:88}, and motional broadening to the anti-Zeno effect.

\begin{figure}
    \begin{center}
    \includegraphics[width=8cm]{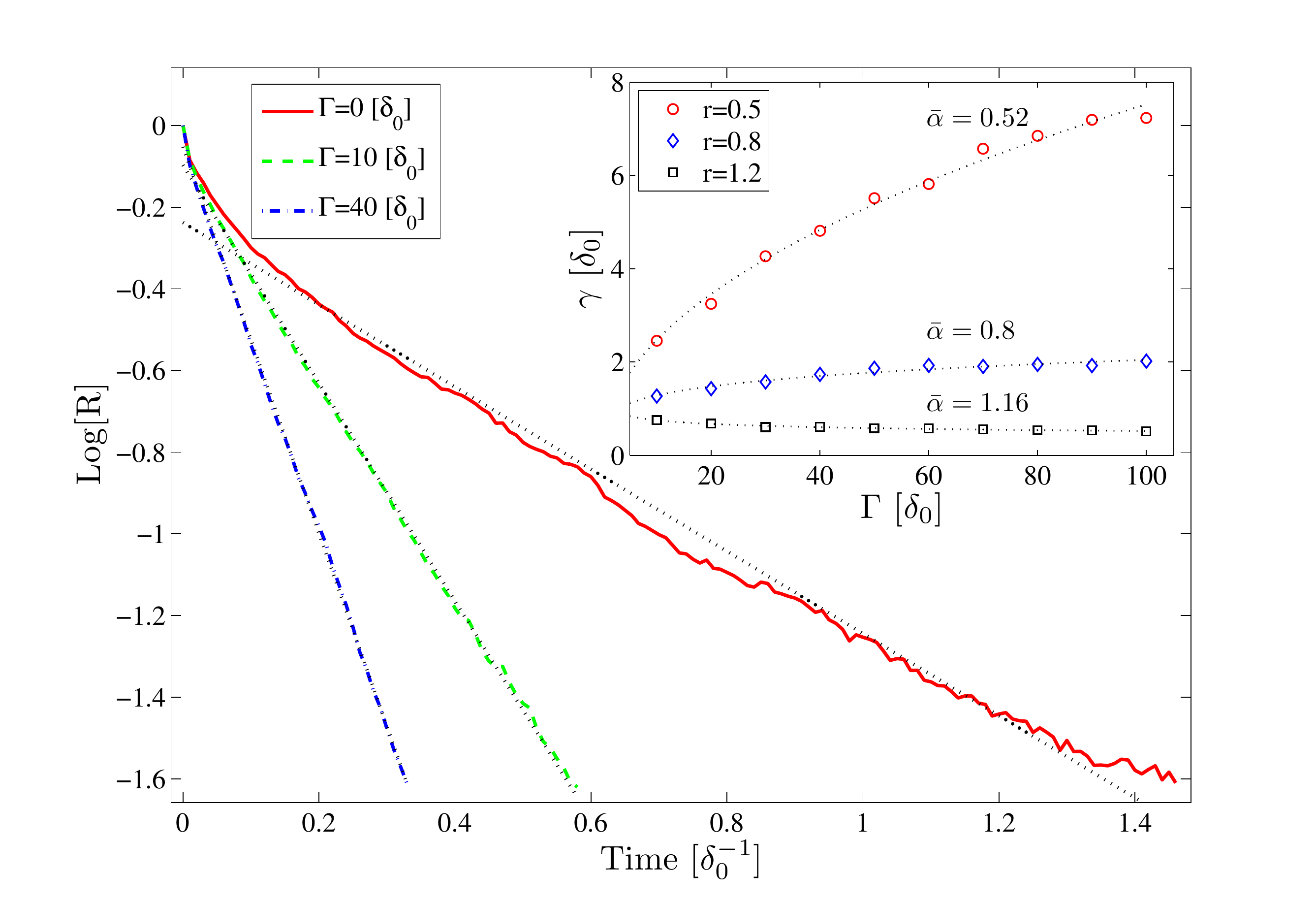}
    \end{center}\caption{Numerical simulation of the logarithm of the coherence using an ensemble of 10000 particles following the Poisson discrete fluctuations model with three different rates $\Gamma$ and detunings following a Student's t-distribution with $r=0.5$. The dotted (black) lines are linear fits, validating the prediction of \eqref{new_width_scaling_with_Gamma_eq} that as $\Gamma$ increases the decay becomes exponential. We extract the decay rate, $\gamma$, by fitting the simulated coherence for $T>10\Gamma^{-1}$ to an exponentially decaying function $A e^{-\gamma T}$. The inset shows $\gamma$ as a function of $\Gamma$ for different values of the distribution parameter $r$. The dotted lines are fits to the function $\gamma=a\Gamma^{1-\bar{\alpha}}+b$ (which has the functional form derived in \eqref{new_width_scaling_with_Gamma_eq}. The extracted exponents agree well with the expected values $\bar{\alpha}=r$.}\label{numerical_study_of_the_width}
\end{figure}

\emph{Asymptotic lineshape}.---We rewrite \eqref{coherence_stable_with_col} using the typical time between collisions $\Delta t_j\sim \Gamma^{-1}$, and the inhomogeneous decay rate $\gamma_0=c_\alpha^{1/\alpha}$, and obtain in the limit of many collisions $\Gamma T\gg 1$: $R(T)\approx e^{-\gamma_0^\alpha \Gamma^{1-\alpha} T}$. The asymptotic behavior of the coherence decays exponentially with time, and the decay rate is given by
\begin{equation}\label{new_width_scaling_with_Gamma_eq}
\gamma=\gamma_0^\alpha \Gamma^{1-\alpha} \ \ .
\end{equation}
This equation for $\alpha=2$ was verified experimentally with optically-trapped cold atomic ensemble \cite{PhysRevLett.105.093001}. Since \eqref{coherence_stable_with_col} is true only for a stable distribution, we test the validity of \eqref{new_width_scaling_with_Gamma_eq} for distributions in the domain of attraction of an $\alpha$-stable distribution in numerical simulations. The results of these simulations for a Student's t-distribution are plotted in Fig. \ref{numerical_study_of_the_width}, and show that as the fluctuation rate increases the decay of coherence indeed becomes exponential. From the numerical curves we extract the decay rate and plot it in the inset of Fig. \ref{numerical_study_of_the_width} for various values of the distribution parameter $r$ and $\Gamma$. The functional form of the decay rates confirms the prediction of \eqref{new_width_scaling_with_Gamma_eq} with the correct $\alpha$.

\begin{figure}
    \begin{center}
    \includegraphics[width=8cm]{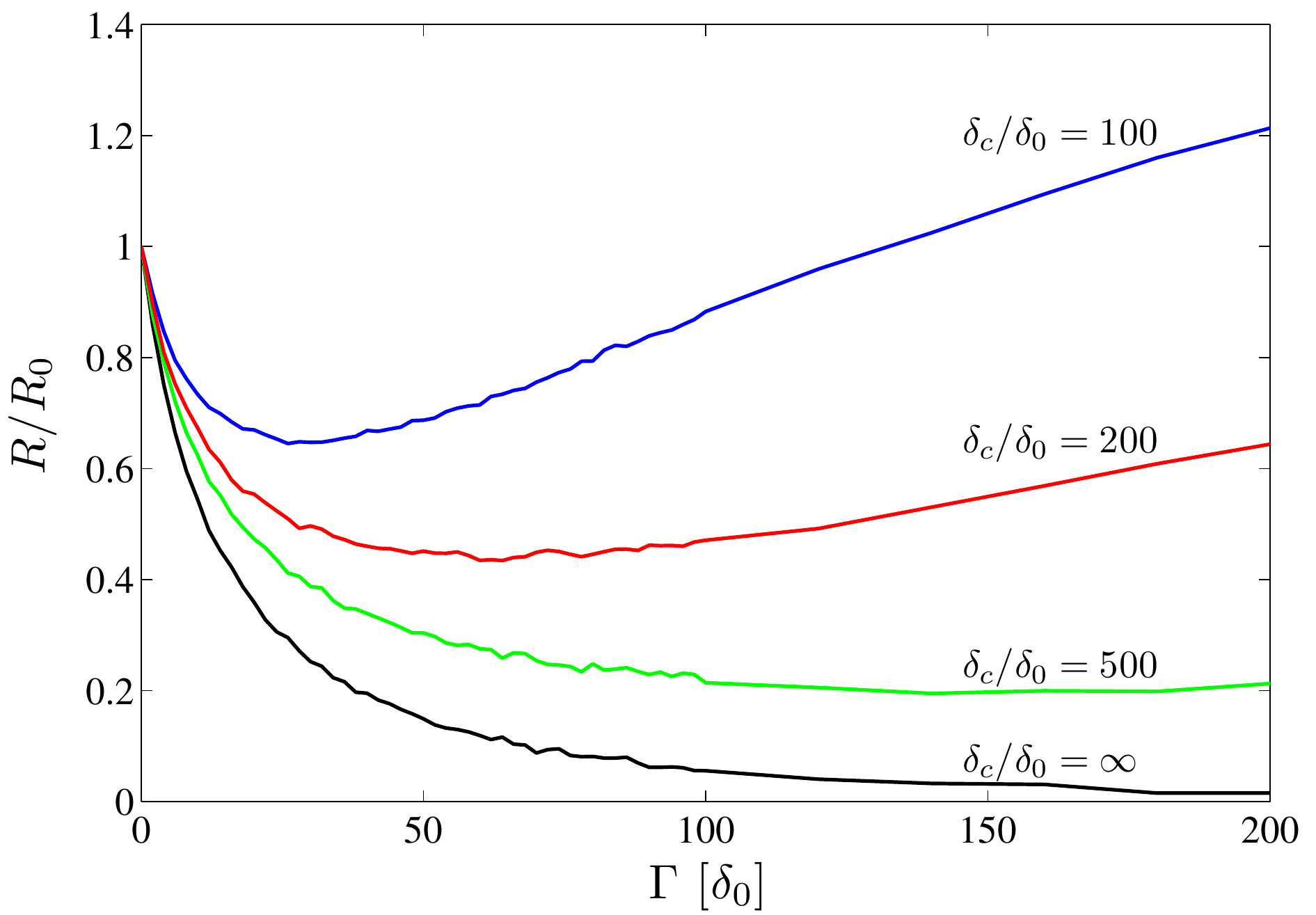}
    \end{center}\caption{The effect of a cutoff in the detuning distribution on the motional broadening phenomenon. The figure shows numerical simulation of the normalized coherence $R(T)/R_0(T)$ at time $T=0.5\delta_0^{-1}$ versus the fluctuations rate $\Gamma$. The detuning distribution is taken to be the Student's t-distribution [see \eqref{students_t_distribution}] with $r=0.5$, and the fluctuations are Poissonian. Each curve is calculated for $P_0$ truncated at a cutoff detuning $\delta_c$.  As expected in motional broadenings, the coherence is decreasing for small $\Gamma$, and only for higher values the behavior is changing to motional narrowing.}\label{effect_of_cutoff_figure}
\end{figure}

\emph{The effect of a cutoff}.---In real physical situations the detuning distribution can not have a diverging first moment, and the heavy tail scaling can be sustained up to some cutoff $\delta_c$. For $P_0(\delta)$ with a characteristic exponent $\alpha<1$, the order of magnitude of the sum $\sum_{j=1}^n \Delta t_j \delta_j$ is the same as of $\text{max}\{t_j \delta_j \}$. The effect of the cutoff is therefore negligible as long as $\text{Prob}(\text{max}[\{\delta_j\}]>\delta_c)\ll 1$  (we assume the average $\langle \Delta t_j \rangle=\Gamma^{-1}<\infty$). This probability depends on the number of collisions, which is roughly given by $\Gamma T$. An estimate of this probability yields that the effect of the cutoff is insignificant for $\Gamma T\ll(\delta_c/\delta_0)^\alpha$, where $\delta_0$ is the typical scale of the detuning distribution [see \eqref{students_t_distribution}]. This means that for a given observation time $T$, motional broadening persists up to fluctuation rate on the order of $(\delta_c/\delta_0)^\alpha T^{-1}$. This qualitative picture is demonstrated in numerical simulations plotted in Fig. \ref{effect_of_cutoff_figure}. Motional broadening prevails for small $\Gamma$, later changing to motional narrowing once the cutoff is sampled and discovered.

\emph{The analogy to diffusion in real space}.---The TLS ensemble coherence problem can be mapped to that of particles performing diffusion in real-space, where the detuning rate and accumulated phase are mapped to velocity and position, respectively. In this analogy, the diffusion problem assumes an ensemble of particles with a steady-state distribution of velocities, starting all from the same point in space. In the absence of collisions, the particles are ballistically expanding and the width of their position distribution grows linearly with time. With collisions, our criterion yields that the width of the particles' spatial distribution grows faster than ballistic (super-ballistic) for heavy-tailed velocity distribution. Super-ballistic diffusion is known to exist in turbulent flow \cite{PhysRevLett.58.1100}. In the context of spatial diffusion, a particularly interesting implementation of the model considered in this paper can be achieved. It was shown that the steady-state velocity distribution of atoms in a polarization lattice follows a power law with an exponent that depends on the lattice depth \cite{limits_sisyphus_1991}. As a result, in such a system the diffusion becomes anomalous \cite{Marksteiner1996}. Going to a low enough lattice depth, it may be possible to observe motional broadening, namely diffusion whose scaling with time is faster than ballistic.

\section{Conclusions}\label{conclusions_section}

In conclusion, inhomogeneous broadened spectrum of an ensemble of two-level systems is modified when the energy of each system is fluctuating. We have shown that although most commonly the fluctuations lead to narrowing of the spectrum, under certain conditions the opposite can occur. This crossover is determined by the tail behavior of the inhomogeneous energy distribution, with motional broadening arising for distributions of divergent first moment. Finally we have drawn the analogy to the anti-Zeno effect and to faster-than-ballistic spatial diffusion.

We thank Shlomi Kotler for helpful discussions. Michael Aizenman thanks the Department of Physics of Complex Systems and the Mathematics Department at the Weizmann Institute of Science for hospitality at a visit during which some of the work was done. This work was partially supported by MIDAS, MINERVA, ISF, DIP, NSF grant DMS-0602360, and BSF grant 710021.

\end{document}